# Study of ARPES data and d-wave superconductivity using electronic models in two dimensions

A. Moreo, A. Nazarenko, S. Haas, A. Sandvik, and E. Dagotto

*Department of Physics and National High Magnetic Field Lab, Florida State University, Tallahassee, FL 32306*

**Abstract**

We review the results of an extensive investigation of photoemission spectral weight using electronic models for the high-Tc superconductors. Here we show that some recently reported unusual features of the cuprates namely the presence of (i) flat bands, (ii) antiferromagnetically induced weight, and (iii) small quasiparticle bandwidths have all a natural explanation within the context of holes moving in the presence of robust antiferromagnetic correlations. Introducing interactions among the hole carriers, a model is constructed which has $d_{x^2-y^2}$ superconductivity, an optimal doping of $\sim 15\%$ (caused by the presence of a large density of states at the top of the valence band), and a critical temperature $\sim 100K$.

In the last couple of years, angle-resolved photoemission (ARPES) data introduced important constraints on possible theories of high-Tc. These experiments have motivated a considerable body of work by the authors and several collaborators that is here briefly reviewed. We have focussed on three recent experiments on the cuprates, discussed elsewhere in this volume, that reported the following information: (i) Hole-doped cuprates Bi2212, Bi2201, Y123 and Y124 near optimal doping present a universal flat band dispersion near $(\pi, 0)$[1] that cannot be explained using band structure calculations. (ii) Data gathered with a novel ARPES technique have produced indications of antiferromagnetically induced photoemission bands in Bi2212[2]. (iii) The quasiparticle bandwidth of several cuprates at finite doping[1], as well as for the insulator $Sr_2CuO_2Cl_2$[3], is of order of the Heisenberg exchange J. Before the theoretical analysis, let us remind the reader that just a few years ago the early papers on ARPES applied to high-Tc had the tendency to claim in their conclusions that a good agreement between experiments and the band structure calculations was obtained. However, the new ARPES results show that such a conclusion was premature and the influence of strong correlations is *crucial* to explain the data. These are certainly very exciting times for "aficionados" of electronic models for the cuprates!

In recent months, we have documented in the literature the results of an extensive investigation showing that the three experimental features mentioned above can have a natural explanation within the context of simple electronic models for the $CuO_2$ planes, like the 2D $t-J$ model.[4–7] In addition, we have shown that under mild assumptions, even a d-wave superconducting phase with a critical temperature of about 100K at an optimal doping of 15% can also be naturally obtained using these ideas.[8] We urge the reader to consult the original literature for details since in the present short review only a sketch of the main results will be presented. Note that we can only provide some of the relevant references. The rest are given in the papers quoted in this short article.

To start the analysis, let us consider the properties of hole carriers in an antiferromagnet. In Fig.1 we show the dispersion of a hole in a fluctuating Néel background calculated numerically using the two dimensional $t-J$ model on large clusters.[4] We observed that the data can be fitted very well by a dispersion where holes effectively move within their own sublattice to minimize disturbing the antiferromagnetic background. The dispersion has interesting features: (i) the bandwidth (W) is small, proportional to the exchange $J = 0.125$eV, a fact observed since the early studies of strongly coupled electrons using numerical techniques[9]. A vast literature supports the result that $W \sim 2J$. Recently, Quantum Monte Carlo (QMC) simulations of the one band Hubbard model in strong coupling arrived to similar conclusions although their reported bandwidth scales like $W \sim 4J$[10]; (ii) the region around $(\pi, 0)$ is anomalously flat. (actually, plotting the quasiparticle energy together with the experimental results of Ref.[1] a good experiment-theory agreement is obtained[4,11]). QMC simulations show a similar result[11]; (iii) the density of states (DOS) has a large peak at the bottom of the hole band (top of the valence band in the electron language) that will be used below to boost the critical temperature of a gas of quasiparticles once interactions are included.

The numerical value for the quasiparticle bandwidth of this model agrees well with ARPES results both for optimaly doped Bi2212 and also the novel insulator compound $Sr_2CuO_2Cl_2$. This provides strong support to models of correlated electrons of the $t-J$ model family. Even some details of the dispersion match the experiments, as exemplified by the line from $(0,0)$ to $(\pi,\pi)$. On the other hand, theory and experiment disagree for the insulator $Sr_2CuO_2Cl_2$ along the line $(\pi,0)$–$(0,\pi)$ since the $t-J$ model predicts an anomalously small dispersion in this particular direction. However, some of us in collaboration with K. Vos and R. J. Gooding have shown in a recent paper[5] that the addition of a small hopping amplitude along the diagonals of the plaquettes on the 2D square lattice greatly improves the agreement theory-experiment for all momenta including the $(\pi,0)$–$(0,\pi)$ line . The addition of this new parameter in the Hamiltonian, while not quite satisfactory since its microscopic origin is unknown, is not forbidden by symmetry, and thus there is no reason to believe that it cannot be present in the cuprates. The dispersion is shown in Fig.2. It is also important to remark that an even better agreement between theory and experiment was reported in Ref.[5] once a more sophisticated calculation is carried out using the more general three band Hubbard model. Thus, the partial conclusion of this analysis is that the bandwidth of the quasiparticles in the cuprates is likely described by models of strongly correlated electrons.

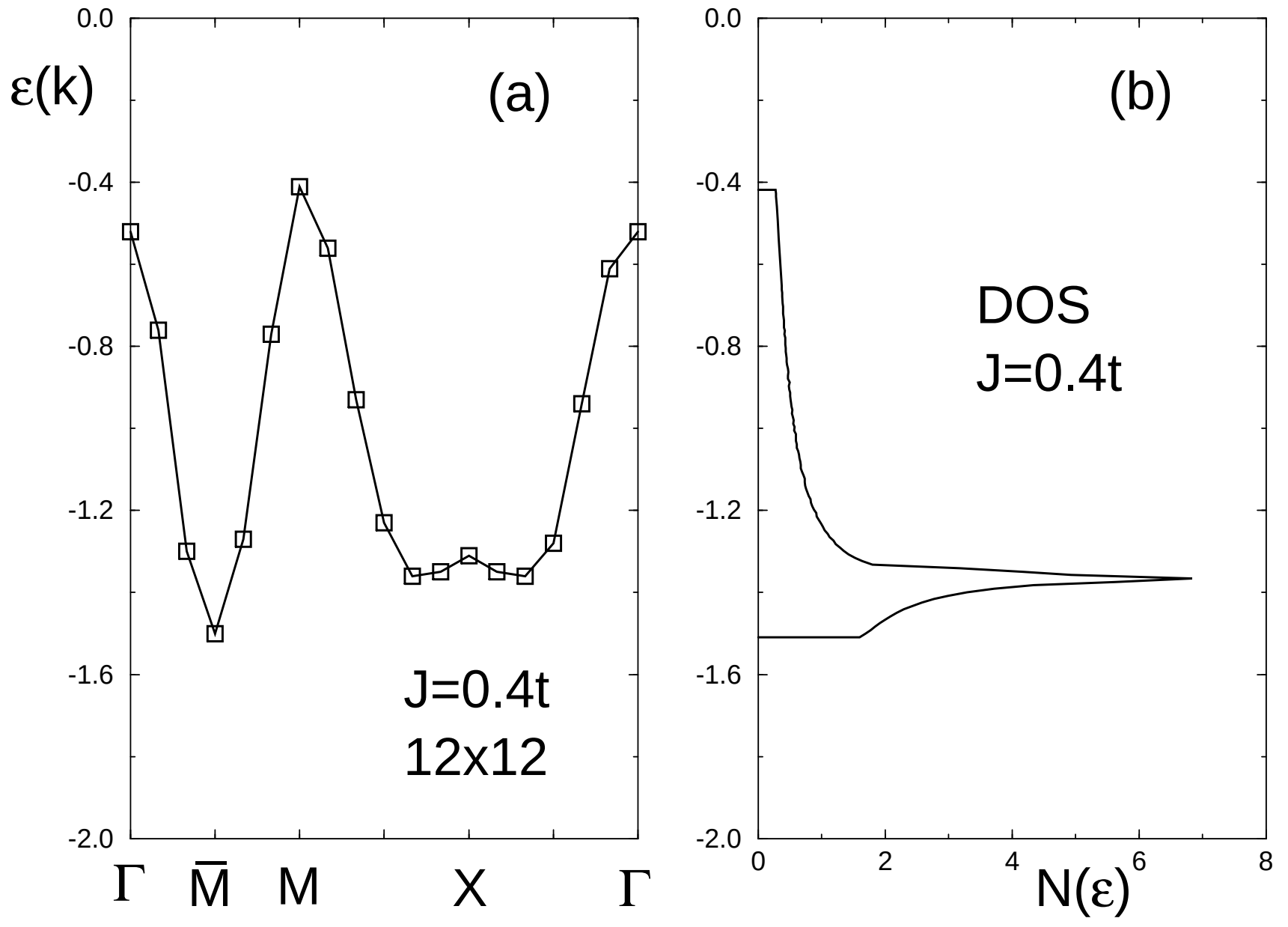

Fig. 1. (a) Energy of a hole in the t − J model, $\epsilon(\mathbf{k})$, vs momentum obtained with a Monte Carlo method on a $12 \times 12$ lattice and $J/t = 0.4$ (in units of t). (b) Density of states obtained from our fit of the numerical data Fig.1a showing the van-Hove singularity between $\bar{M}$ and X. The unit of energy is t (from Ref.[4]).

It is interesting to note that results similar to those obtained for the 2D t − J model at half-filling can also be obtained using the 2D one band Hubbard model. In Fig.3 we show results of a Monte Carlo simulation of this model on clusters of 64 sites. The ARPES intensity is obtained with the maximum entropy technique. While this last technique is crude, it seems to provide results in agreement with the more reliable Exact Diagonalization (ED) analysis of the t − J model. The Hubbard model results of Fig.3 show that working at large U/t and concentrating near the chemical potential, a band disperses with a bandwidth of order 2J, in excellent agreement with old predictions for the t − J model.[9] The rest of the weight is incoherent and appears at large energies of the order of the hopping t. We noticed that small variations of the maximum entropy method may induce a mixing of both bands into a single feature, which produces an incorrect larger bandwidth. Temperature effects are also crucial to obtain the proper results in these studies. Thus, we highly advice to be cautious in the use of maximum entropy techniques for correlated electrons. To test their accuracy, the results must necessarily be contrasted against other techniques that provide dynamical information for the $t-J$ model.

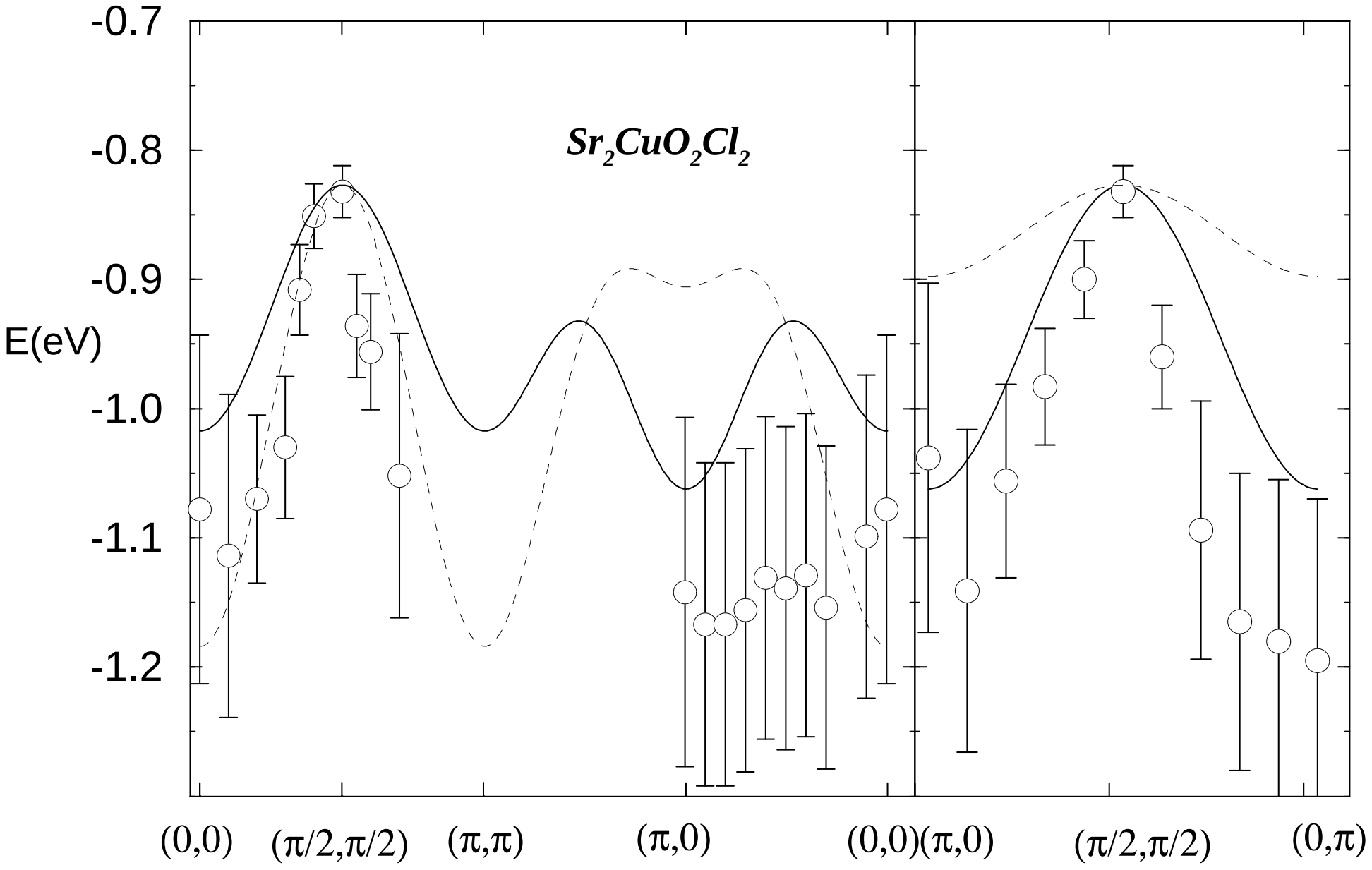

Fig. 2. Quasiparticle dispersion of the $t-t'-J$ model (Ref.[5]) calculated in the Born approximation for an infinite lattice using $t' = -0.35t$ (solid line), $J/t = 0.3$ and $J = 0.125eV$, compared against the experimental ARPES data (open circles) for $Sr_2CuO_2Cl_2$ of Ref.[3]. The dashed line is the $t-J$ model result of Ref.[12].

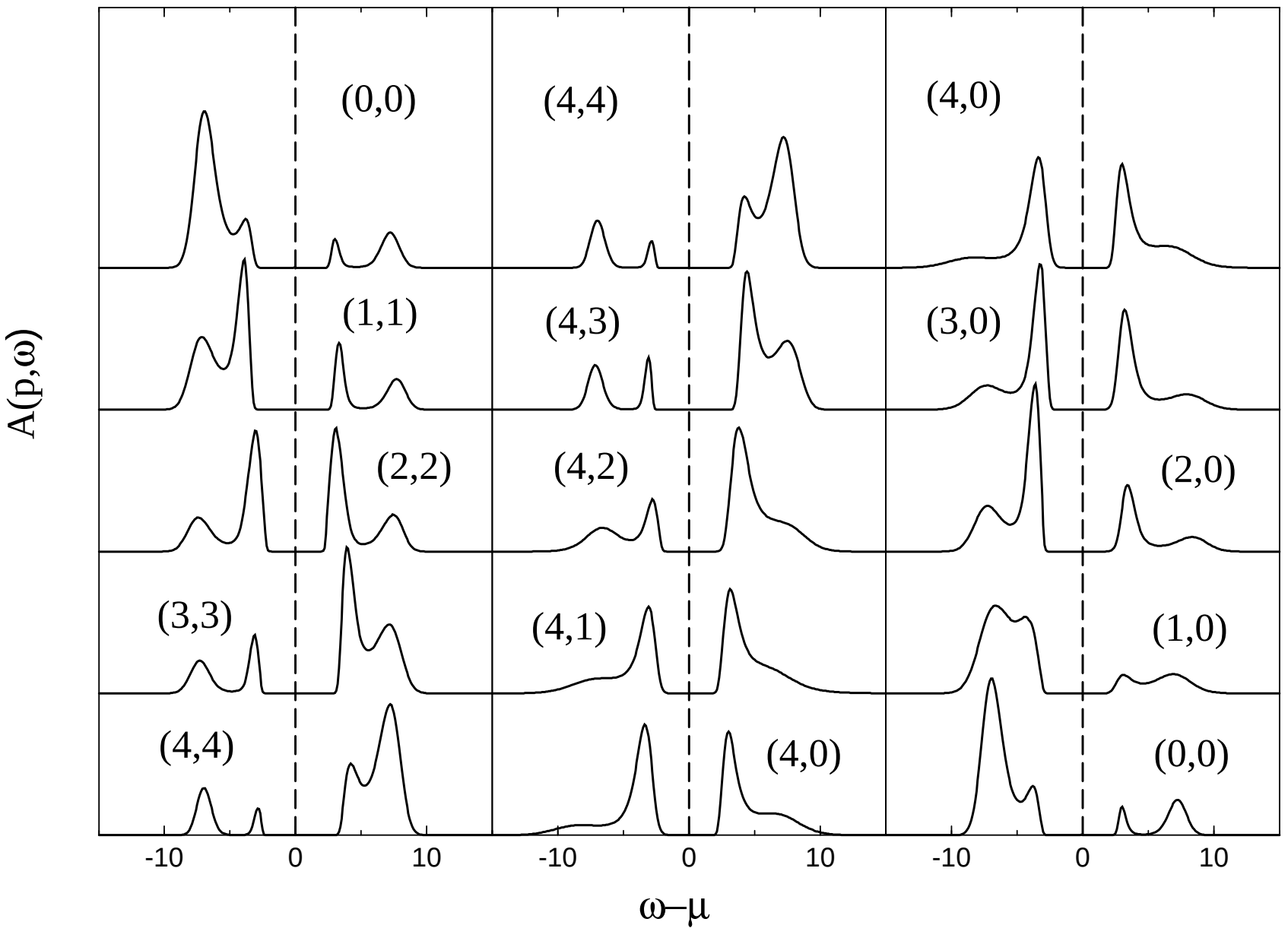

Fig. 3. Spectral weight $A(\mathbf{p}, \omega)$ of the 2D Hubbard model at half-filling obtained with the QMC method supplemented by Maximum-Entropy, on an $8 \times 8$ cluster, $U/t = 10$, and $T = t/4$. (from Ref.[6]).

Before discussing the influence of doping on our results, we want to remark to the reader an interesting prediction of the $t-J$ model namely the presence of "string states" in the ARPES photoemission data. These states have the fol-lowing origin: suppose a hole is injected in an antiferromagnet, as it occurs in the sudden approximation description of a photoemission experiment. When the hole attempts to move, which occurs by moving a spin from a nearest neighbor site $j$ to the original position of the hole $i$, energy is lost since now the spin at site $i$ is aligned *ferromagnetically* with its neighbors (assuming for simplicity a perfect Néel background). As the hole continues its excursion, say along one axis, more and more spins have ferromagnetic links with some of its neighbors. Actually, the penalization in energy grows proportional to the length of the path followed by the hole. Then, to a good approximation holes are confined into a linear potential caused by the spin background. The solution of the Schrödinger equation for a hole in an effective linear potential has several bound states[13]. The lowest energy state corresponds to the dominant peak in the dispersion (that acquires mobility thanks to the antiferromagnetic fluctuations that sometimes break the linear "string" caused by the hole in its movement), while the excited states are precisely the so-called "string states". These ideas have been discussed since the early studies for correlated electrons using the $\mathrm{t-J}$ model and developed by several groups. For details and references see Ref.[9]. Numerical results already in 1990 clearly showed their existence in spite of the strong quantum fluctuations in the ground state.[14] In Fig.4 we show the main quasiparticle peak, as well as one excited state, for the case of the $\mathrm{t-t'-J}$ model in two dimensions using the rainbow approximation[15] (results for the $\mathrm{t-J}$ model are very similar). The first string state beyond the quasiparticle peak is clearly observed at a binding energy of 0.6 eV. However, the reader should notice that the calculation of excited states is somewhat less accurate than the evaluation of the properties of the dominant quasiparticle peak, and thus, the 0.6 eV prediction may have large error bars. The expected intensity is of about 20% of the main peak. These states *should* be present in $Sr_2CuO_2Cl_2$ if our analysis based on the $\mathrm{t-J}$ and $\mathrm{t-t'-J}$ models is correct. This is a clear prediction of these models. We also expect that the main candidate momentum for such an investigation should be $(\pi/2, \pi/2)$, i.e. at the top of the valence band. The only problem for the observation of the string states are effects not taken into account in the theoretical analysis, specially the existence of a robust spurious background in photoemission experiments. We nevertheless encourage our experimental colleagues to search for string states in ARPES data.

The results summarized before have shown that simple electronic models for the cuprates may properly explain several features of the real cuprate materials observed in ARPES data. The theoretical results reported above have been carried out at half-filling, i.e. in the proper density regime for $Sr_2CuO_2Cl_2$. However, the comparison between theoretical predictions and data for hole-doped compounds at *optimal* doping successfully carried out in Ref.[4] required the additional assumption that the dispersion does not change drastically with doping between half-filling and the optimal concentration of about ($\sim$ 15%).[16] Such an assumption is reasonable since results for both Bi2212 and $Sr_2CuO_2Cl_2$ have similar small bandwidths of the order of J. On the other hand, they differ in some of their fine details specifically about momentum $(\pi, 0)$, and thus a strict rigid band approximation is likely to be non valid. Thus, the main issue is to what extent the assumption of rigid band behavior is a good approximation to describe the low hole density region of the cuprates. A more accurate first-principles calculation is required to check the quality of this approximation. This issue has been recently addressed using numerical techniques, mainly of the ED family since temperature and sign effects in Monte Carlo determinantal methods prevent us from analyzing the low temperature behavior of the Hubbard model away from half-filling. In Fig.5, results of an ED analysis using clusters of 16 and 18 sites at a density of $\langle n \rangle = 0.88$ using the $\mathrm{t} - \mathrm{J}$ model are reproduced from Ref.[6]. It is clear that the two-features structure observed in Fig.3 is still present. The bandwidth and shape of the quasiparticle feature at energies smaller than the chemical potential remain similar to those observed before at half-filling. Of course, and as we reported in Ref.[6], as the hole density increases, eventually the quasiparticle structure with bandwidth of order J disappears. Its existence seems correlated with the presence of robust antiferromagnetic correlations in the ground state.

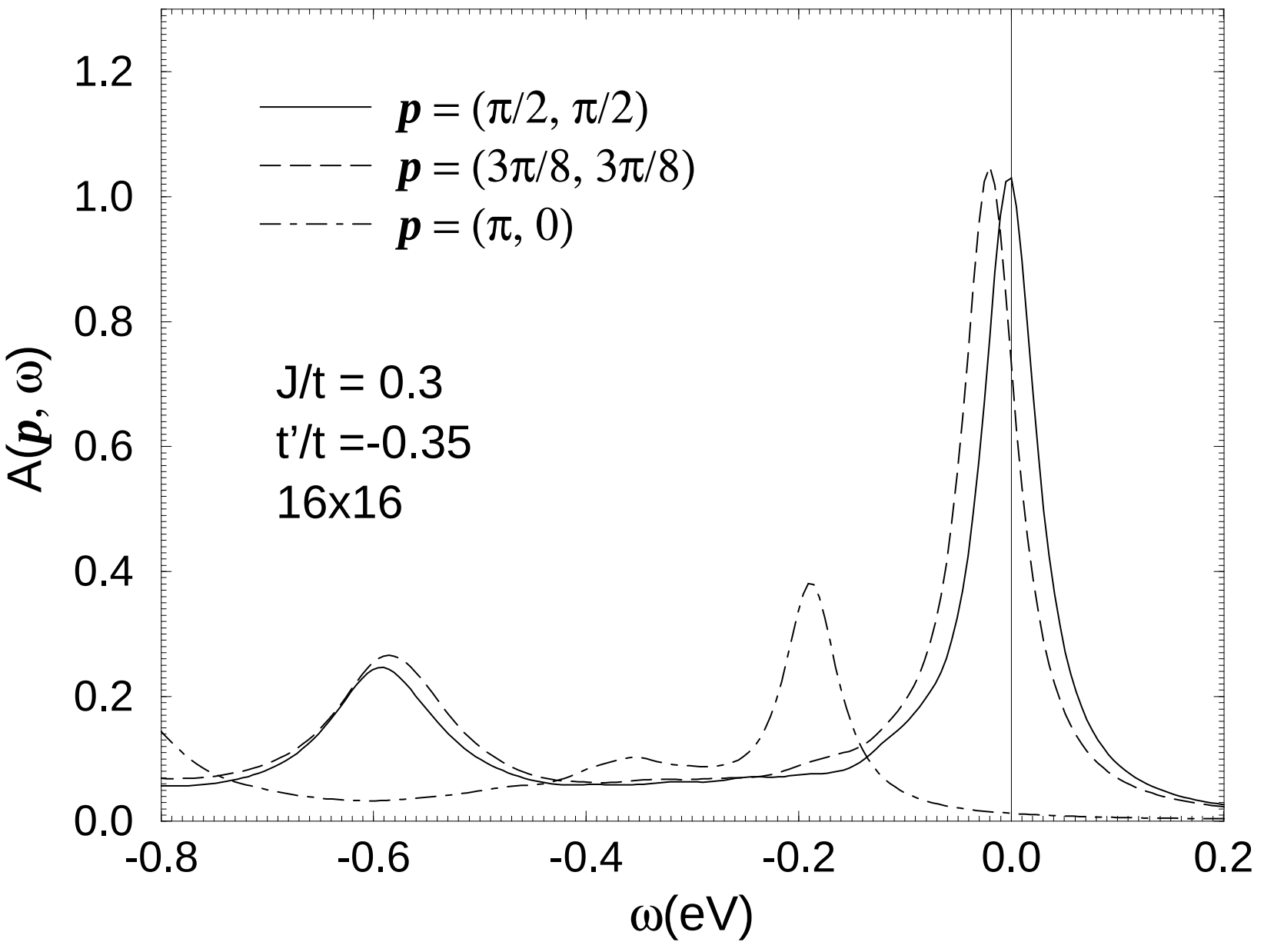

Fig. 4. Spectral weight $A(\mathbf{p},\omega)$ for one hole in an antiferromagnet calculated using the rainbow approximation. The parameters are shown in the figure. The first excited state in the "string picture" is located at $\omega \sim -0.6eV$.

This is a good place to address another consequence of antiferromagnetic correlations in the ground state of models of correlated electrons, namely the creation of new structure in the ARPES signal. We have already seen that it produces a small bandwidth of order J and flat regions near $(\pi, 0)$. However, in the introduction we mentioned a third very interesting recent experiment that reported the presence of antiferromagnetically induced bands in Bi2212.[2] The physics of this effect is as follows. First ,consider the half-filled case where the antiferromagnetic long-range order effectively doubles the size of the unit cell, thus reducing in half the Brillouin zone. This dynamically generated reduction causes the appearance of an extra symmetry in the ARPES results. For example, and as shown in Fig.1, the dispersion along the line from $(0,0)$ to $(\pi,\pi)$ is symmetric with respect to the $(\pi/2,\pi/2)$ point. This means that in the presence of Néel order momenta *above* the Fermi momentum $\mathbf{p}_{\mathrm{F}}$, corresponding to non-interacting electrons for the same density, will show a nonzero signal in an ARPES experiment. The existence of this weight is a direct consequence of strong correlations, and it was discussed some time ago by Kampf and Schrieffer[17] using weak coupling diagrammatic techniques. Motivated by the novel experiments by Aebi et al.[2], we decided to carry out a numerical study of the strength of these bands vs. the antiferromagnetic correlation length $\xi_{AF}$ in the strong coupling regime. One of our results in shown in Fig.6. We show there the spectral function studied on small clusters away from half-filling. It is clearly observed that there is a large peak near the chemical potential even for momenta above $(\pi/2,\pi/2)$ where the Fermi surface for a weakly interacting gas of fermions is. This is the effect predicted by Kampf and Schrieffer, i.e. the presence of a antiferromagnetically induced quasiparticle band in models for the cuprates. Note that the momentum dependence of the results is crucial to distinguish this effect from a simple "lower Hubbard band" formation which would occur even in the atomic limit. For details see Ref.[7].

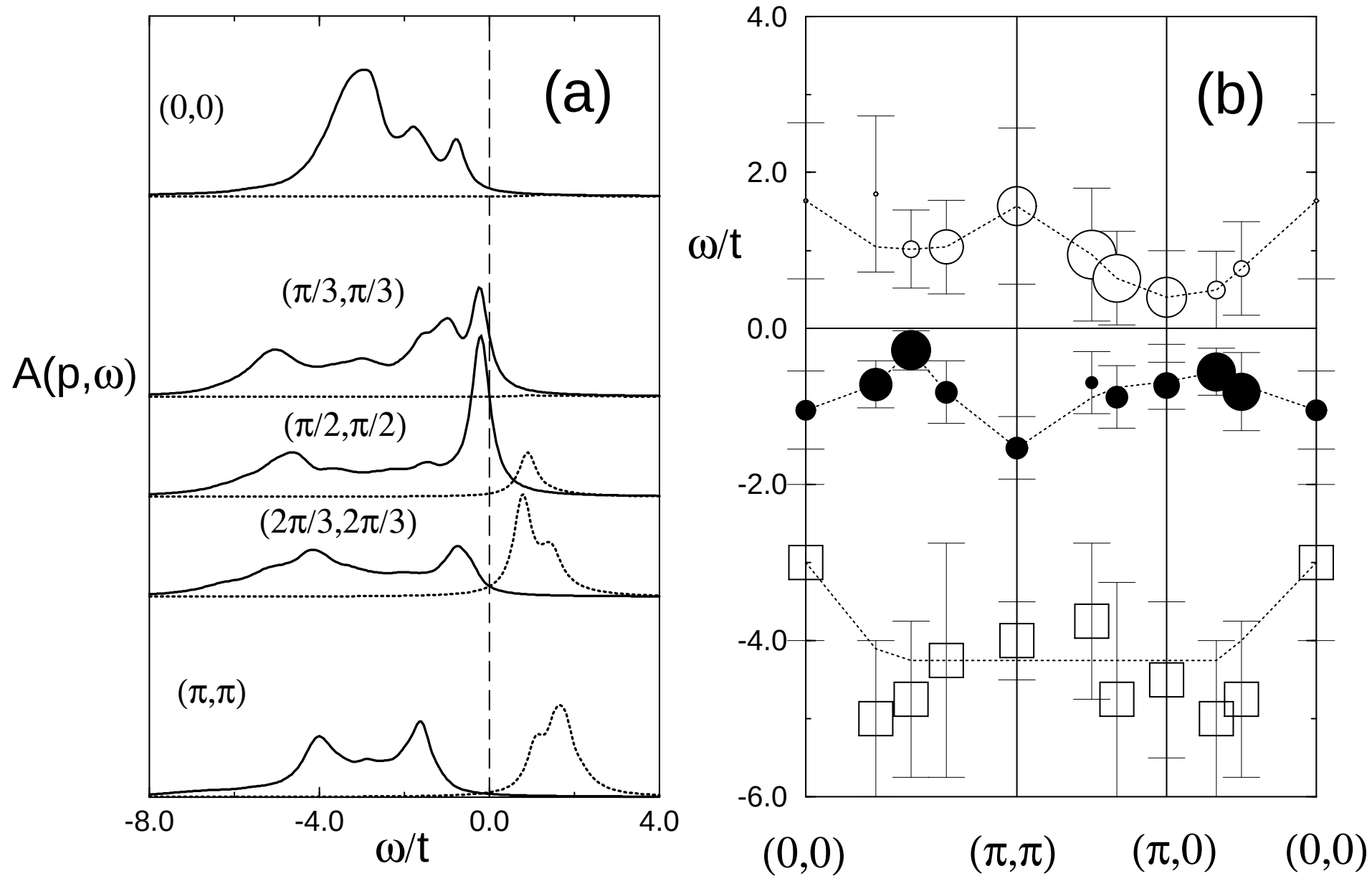

Fig. 5. A$(\mathbf{p},\omega)$ for density $\langle n \rangle \approx 0.88$ (i.e. two holes on the 16 and 18 sites clusters) using the $\mathrm{t}-\mathrm{J}$ model. In (a) the PES intensity ($\omega < 0$) is shown with a solid line, while the IPES intensity ($\omega > 0$)is given by a dotted line. The chemical potential is located at $\omega = 0$. In (b) the full and open circles represent the PES and IPES intensities, respectively, of the peaks the closest to the Fermi energy. Their area is proportional to the intensity. The error bars denote the width of the peak as observed in Fig.5a (sometimes to a given broad peak several poles contribute appreciably). The full squares at $\omega \sim -4t$ represent the center of the broad valence band weight, and the area of the squares is *not* proportional to their intensity.

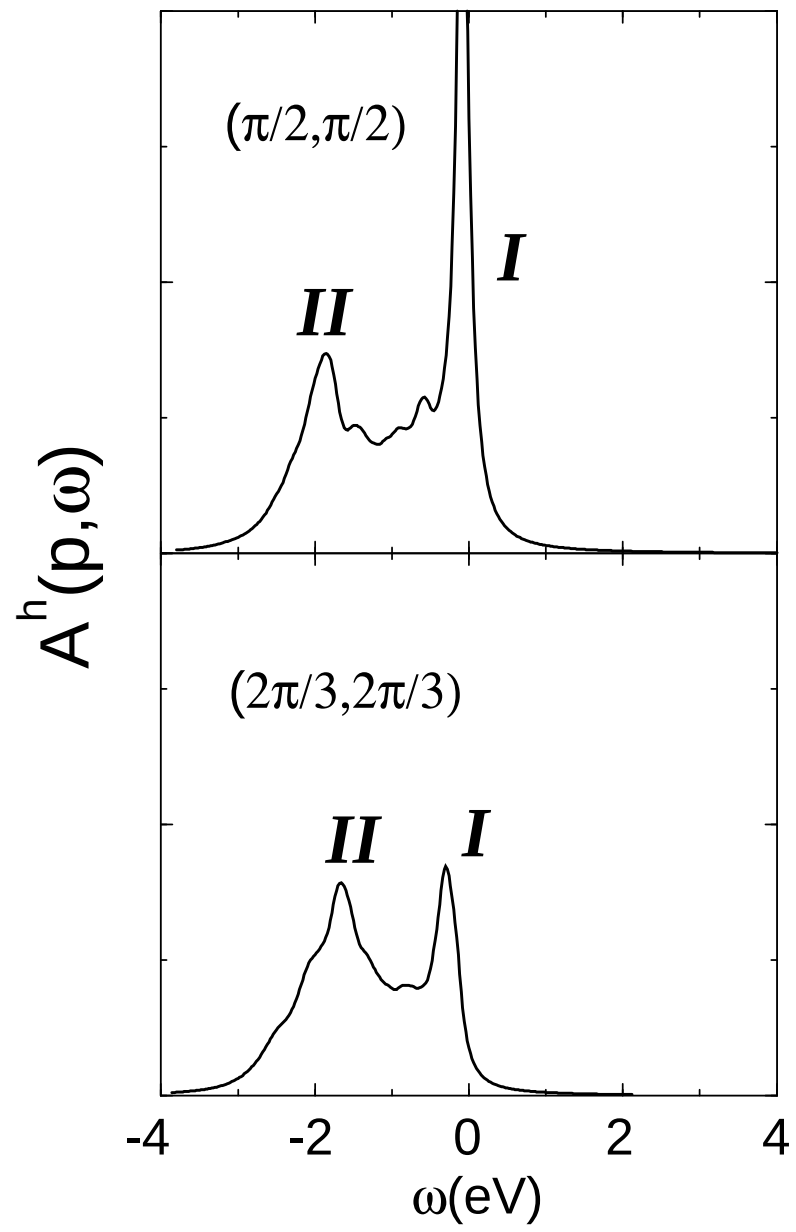

Fig. 6. PES $A(\mathbf{p},\omega)$ (here denoted as $A^h(\mathbf{p},\omega)$) for the $t-J$ model at $\langle n \rangle \sim 0.88$, $J/t = 0.4$, and clusters of 16 and 18 sites. We use $\delta = 0.25t$ and $t = 0.4$eV. I is the quasiparticle-like part and II is the incoherent background.

We have observed that the strength of this effect rapidly dies out away from half-filling, but it can still be "observed" at 15% doping where $\xi_{AF}$ is about a couple of lattice spacings (for the definition of what we label as observable in this context, see Ref.[7]). Then, although we cannot prove that indeed Aebi et al.'s results are evidence of the bands caused by antiferromagnetic correlations, at least our theoretical results and their data are *compatible*, namely of the same order of magnitude.

After this rapid description of our main results for the normal state ARPES signal, we should address the superconducting state. Interesting progress has been recently made in this context.[8] Let us first assume that the normal state of the cuprates can be roughly approximated by a gas of weakly interacting quasiparticles with the dispersion calculated, for example, in Fig.1. This approximation is correct even if there is no long range order, as long as $\xi_{AF}$ is robust enough (i.e. such that a hole feels that it is immersed in a *local* Néel environment). The interaction between holes can also be deduced from the behavior of two holes in an antiferromagnet which has been studied using numerical techniques.[9] The effective interaction amounts to an attraction of strength J which is basically operative over a short distance of roughly one lattice spacing. The study of this model is described elsewhere in this volume by Elbio Dagotto et al. and will not be repeated here. Let us just say that when the chemical potential reaches the large peak in the DOS of Fig.1 at the bottom of the hole band, the critical temperature of the effective model reaches its maximum value. The symmetry of the superconducting order parameter is $d_{x^2-y^2}$ and the Tc is about 100K at a hole doping of 15%.[18] This model, called the "Antiferromagnetic van Hove" (AFVH) model due to its features combining pairing through a magnetic mechanism with the presence of a large DOS causing a large Tc, is very promising, and currently we are actively investigating its properties. It has already been shown that the Hall coefficient[4] and $2\Delta/kT_c \sim 5$[8] are in good agreement with experiments.

Finally, note that our results are based on the assumption that the quasi-particle dispersion does not change much with doping near half-filling. Fig.5 provides some evidence supporting it. The key point we need for the AFVH scenario is that the large peak in the DOS observed at half-filling survives as carriers are introduced. In Fig.7 we show the density of states, N($\omega$), vs $\omega - \mu$ using exact diagonalization techniques on clusters of 16 and 18 sites and two and four holes. These results show that a large peak is still present in the DOS, and this peak moves across the chemical potential as the electronic density is changed, giving more support to the AFVH ideas.

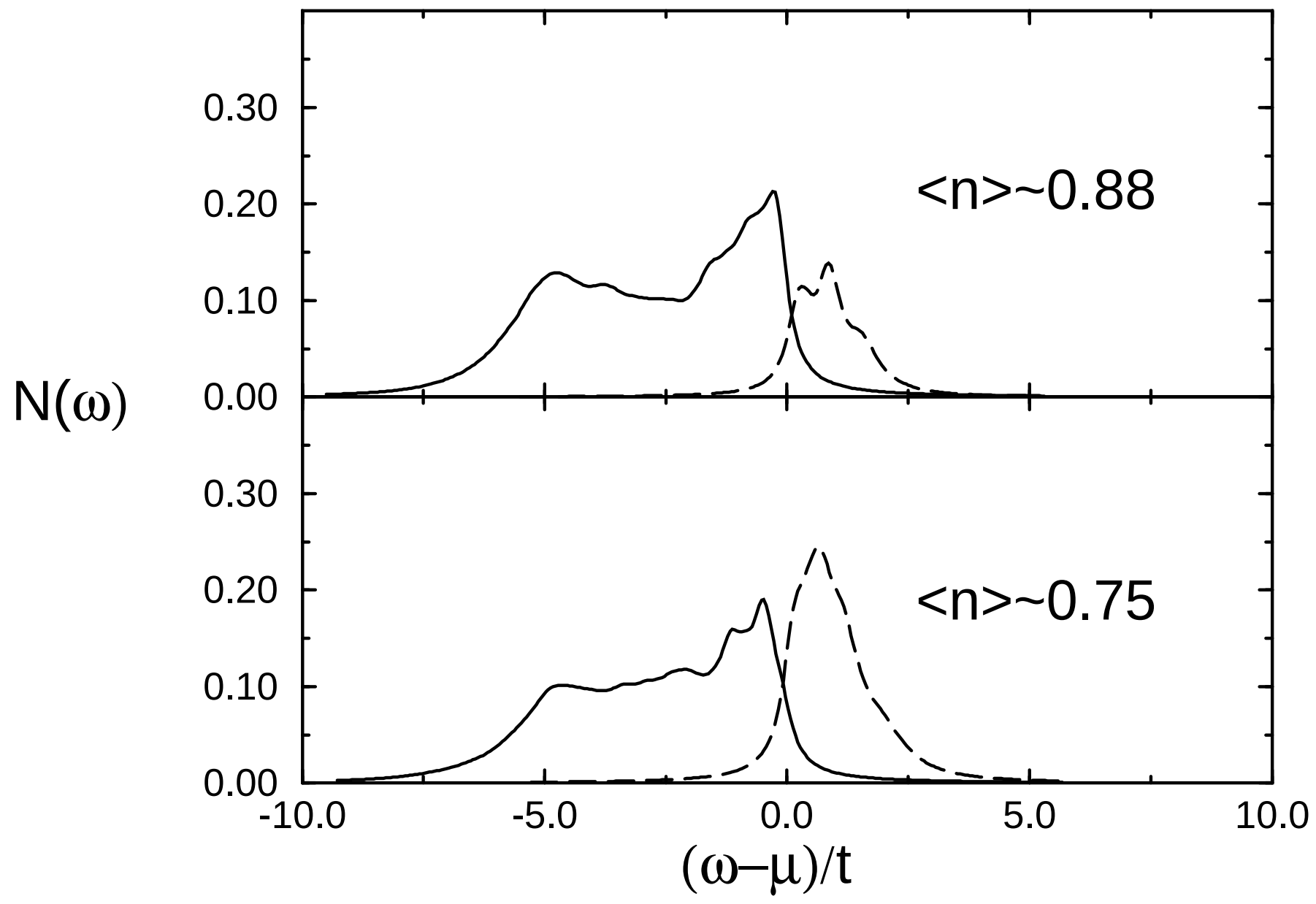

Fig. 7. Density of states of the $t - J$ model obtained with clusters of 16 and 18 sites, at $J/t = 0.4$, and at the electronic densities shown.

Summarizing, recently considerable progress has been made in the understanding of ARPES experimental data using models of correlated electrons. Several unusual features have a natural explanation once strong correlations are properly included in the calculations. Flat bands, antiferromagnetically induced bands, and small bandwidths can be accounted for in this way. These results are difficult to explain with band structure calculations. Even superconductivity in the d-channel is obtained with a Tc of about 100K. We have benefited from conversations with many colleagues, the list being too large to reproduce here. We warmly thank all of them. The authors are supported by the Office of Naval Research under grant ONR N00014-93-0495, the donors of the Petroleum Research Fund administered by the ACS, the National High Magnetic Field Lab, and the Supercomputer Computations Research Institute.